\documentclass[traditabstract]{aa}

\usepackage{amsmath} 
\usepackage{amssymb} 
\usepackage{txfonts}
\usepackage{graphicx}
\usepackage[colorlinks=true,linkcolor=blue,citecolor=blue,urlcolor=blue]{hyperref} 
\usepackage{natbib}
\bibpunct{(}{)}{;}{a}{}{,} 
\pdfoutput=1

\begin{document}

\title{Analytical expressions for the deprojected S\'ersic model}

\author{M. Baes \and G. Gentile}

\institute{%
  Sterrenkundig Observatorium, Universiteit Gent, Krijgslaan
  281-S9, B-9000 Gent, Belgium }

\date{\centering Received 8 Sep 2010 / Accepted 23 Sep 2010}
 
\abstract{The S\'ersic model has become the standard to parametrize
  the surface brightness distribution of early-type galaxies and
  bulges of spiral galaxies. A major problem is that the deprojection
  of the S\'ersic surface brightness profile to a luminosity density
  cannot be executed analytically for general values of the S\'ersic
  index. Mazure \& Capelato (2002) used the Mathematica$^\circledR$
  computer package to derive an expression of the S\'ersic luminosity
  density in terms of the Meijer $G$ function for integer values of
  the S\'ersic index. We generalize this work using analytical means
  and use Mellin integral transforms to derive an exact, analytical
  expression for the luminosity density in terms of the Fox $H$
  function for {\em{all}} values of the S\'ersic index. We derive
  simplified expressions for the luminosity density, cumulative
  luminosity and gravitational potential in terms of the Meijer $G$
  function for all rational values of the S\'ersic index and we
  investigate their asymptotic behaviour at small and large radii. As
  implementations of the Meijer $G$ function are nowadays available
  both in symbolic computer algebra packages and as high-performance
  computing code, our results open up the possibility to calculate the
  density of the S\'ersic models to arbitrary precision. }


\maketitle

\section{Introduction}

The S\'ersic model \citep{1968adga.book.....S} is a three-parameter
model for the surface brightness profile of galaxies that has been
introduced as a generalization of the de Vaucouleurs $R^{1/4}$ model
\citep{1948AnAp...11..247D}. It is defined as
\begin{equation}
  I(R)
  =
  I_0
  \exp\left[-b\left(\frac{R}{R_{\text{e}}}\right)^{1/m}\right]
\label{SersicI}
\end{equation}
where $I(R)$ is the surface brightness at radius (on the plane of the
sky) $R$, $I_0$ is the central surface brightness, $R_{\text{e}}$ is
the effective radius, and $m$ is the so-called S\'ersic index that
describes the index of the logarithmic slope power law. The parameter
$b$ is a dimensionless parameter that depends on the S\'ersic index
$m$ and whose value can be derived from the requirement that the
isophote corresponding to $R_{\text{e}}$ encloses half of the total
flux.

Over the past two decades, the S\'ersic model has become the standard
to describe the surface brightness profiles of early-type galaxies and
bulges of spiral galaxies \citep[e.g.][]{1988MNRAS.232..239D,
  1993MNRAS.265.1013C, 1994MNRAS.271..523D, 1994ApJS...93..397C,
  1995MNRAS.275..874A, 1997A&A...321..111P, 2001A&A...368...16M,
  2003AJ....125.2936G, 2006MNRAS.371....2A, 2009MNRAS.393.1531G}. In
the past few years, S\'ersic-like models have also gained popularity
as a model to describe the spherically averaged profiles for dark
matter haloes. While models with a power law behaviour at small and
large radii were preferred in earlier simulations
\citep[e.g.][]{1997ApJ...490..493N, 1999MNRAS.310.1147M}, other models
seem to fit the mass density distribution of more recent (and higher
resolution) simulations better \citep{2004MNRAS.349.1039N,
  2010MNRAS.402...21N, 2005ApJ...624L..85M, 2006AJ....132.2685M,
  2006AJ....132.2701G, 2006MNRAS.373..632A, 2008MNRAS.387..536G,
  2008MNRAS.390L..64D, 2009MNRAS.398L..21S}. Among these models, also
the S\'ersic model (i.e.\ a model where the projected surface density
is described by a S\'ersic law) has been proposed as a universal
description for simulated dark matter haloes.

Mainly as a result of its popularity to describe the surface
brightness profiles of early-type galaxies, the properties of the
S\'ersic model have been examined in great
detail. \citet{1991A&A...249...99C} and \citet{1997A&A...321..724C}
give a detailed description of the properties of the S\'ersic model,
including spatial and dynamical
properties. \citet{1999A&A...352..447C} give a full asymptotic
expansion of the dimensionless scale factor $b$ of the S\'ersic
model. \citet{2005PASA...22..118G} present a compendium of
mathematical formulae on photometric parameters such as Kron magnitues
and Petrosian indices, and \citet{2004A&A...415..839C} and
\citet{2007JCAP...07..006E} investigate the lensing properties. A
major problem with the S\'ersic models is that the deprojection of the
surface brightness profile to a luminosity density is in general
non-analytical. In practice, one often uses approximations for the
S\'ersic models when the luminosity density (or the mass density when
the S\'ersic model is used to describe the distribution of dark
matter) is necessary \citep[e.g.][]{1997A&A...321..111P,
  1999MNRAS.309..481L, 2002MNRAS.333..510T}. An unexpected analytical
progress was the work by \citet{2002A&A...383..384M}, who demonstrated
that it is possible to elegantly write the spatial luminosity density
of the S\'ersic model in terms of the Meijer $G$
function. Unfortunately, their result only holds for integer values of
the S\'ersic index, which is a significant limitation for practical
applications. Moreover, since their result fell as a {\em{deus ex
    machina}} out of the Mathematica$^\circledR$ computer algebra
package, it is hard to see whether it can be generalized to all
S\'ersic indices.

In this paper, we tackle the deprojection of the S\'ersic surface
brightness profile using analytical means. We apply an integration
method based on Mellin integral transforms and derive an exact,
analytical expression for the luminosity density in terms of the Fox
$H$ function for {\em{all}} values of the S\'ersic index $m$. For
rational values of $m$, the luminosity density can be written in terms
of the Meijer $G$ function. As the Meijer $G$ function is nowadays
available both in symbolic computer algebra packages and as
high-performance computing code, this opens up the possibility to
calculate the luminosity density of the S\'ersic models to arbitrary
precision. The wide range of analytical properties of the Meijer $G$
function also allows to easily study the asymptotic behaviour at small
and large radii and to compute derivative quantities such as the
cumulative luminosity and the gravitational potential analytically.

This paper is organized as follows. In Section~{\ref{Density.sec}} we
discuss the general deprojection of the S\'ersic surface brightness
profile using the Mellin transform method and we present a general
expression in terms of the Fox $H$ function. In
Section~{\ref{DensityG.sec}} we present more simple expressions for
integer, half-integer and rational values of $m$ in terms of the
Meijer $G$ function and discuss two special, interesting cases. In
Section~{\ref{totalluminosity.sec}} we use the expressions for the
luminosity density to calculate the total luminosity of the S\'ersic
models, which serves as a sanity check on the derived formulae. In
Section~{\ref{Asymptotic.sec}} we investigate the asymptotic behaviour
of the luminosity density and in Section~{\ref{Derived.sec}} we
derive analytical expressions for the cumulative luminosity and the
gravitational potential. Finally, in Appendix~{\ref{Meijer.sec}} we
introduce the Meijer $G$ and Fox $H$ functions and discuss some of
their most useful properties.

\section{The luminosity density of the S\'ersic model as a Fox $H$
  function}
\label{Density.sec}

In spherical symmetry, the deprojected luminosity density $\nu(r)$ at
the spatial radius $r$, can be found from the surface brightness
profile $I(R)$ through the standard deprojection formula,
\begin{equation}
  \nu(r)
  =
  -\frac{1}{\pi}
  \int_r^\infty \frac{{\text{d}}I}{{\text{d}}R}(R)\,
  \frac{{\text{d}}R}{\sqrt{R^2-r^2}}
  \label{deprojection}
\end{equation}
If we introduce the reduced radial coordinate
\begin{equation}
  s=\frac{r}{R_{\text{e}}}
\end{equation} 
we find
\begin{equation}
  \nu(r)
  =
  \frac{I_0}{R_{\text{e}}}\,
  \frac{b}{m\,\pi}\,
  \int_s^\infty
  \frac{{\text{e}}^{-bt^{1/m}} t^{\frac1m-1}\,{\text{d}}t}
  {\sqrt{t^2-s^2}}
\label{depint}
\end{equation}
The integral (\ref{depint}) cannot be evaluated in terms of elementary
functions or even the standard special functions for general values of
$m$. In order to evaluate it, we use a general method that builds on
Mellin integral transforms and has become known as the Mellin
transform method \citep{Marichev, Adam, Fikioris}. The Mellin
transform ${\mathfrak{M}}_f(u)$ of a function $f(z)$ is defined as
\begin{equation}
 {\mathfrak{M}}_f(u)
  =
  \phi(u)
  =
  \int_0^\infty f(z)\,z^{u-1}\,{\text{d}}z
\end{equation}
The inverse Mellin transform is found as
\begin{equation}
  {\mathfrak{M}}^{-1}_\phi(z)
  =
  f(z)
  =
  \frac{1}{2\pi i}
  \int_{\mathcal{L}} \phi(u)\,z^{-u}\,{\text{d}}u
\end{equation}
where the ${\mathcal{L}}$ is a line integral over a vertical line in
the complex plane. The driving force behind the Mellin transform
method is the Mellin convolution theorem. The Mellin convolution of
two functions $f_1(z)$ and $f_2(z)$ is defined as
\begin{equation}
  (f_1\star f_2)(z)
  =
  \int_0^\infty 
  f_1(t)\,f_2\left(\frac{z}{t}\right)\,
  \frac{{\text{d}}t}{t}
\end{equation}
Similar as for the well-known Fourier transform analogue, the Mellin
convolution theorem states that the Mellin transform of a Mellin
convolution is equal to the products of the Mellin transforms of the
original functions, 
\begin{equation}
  {\mathfrak{M}}_{f_1\star f_2}(u)
  =
  {\mathfrak{M}}_{f_1}(u)\,{\mathfrak{M}}_{f_2}(u)
\end{equation}
Now it can be shown that any definite integral
\begin{equation}
  f(z)
  =
  \int_0^\infty g(t,z)\,{\text{d}}t
\label{Marichev}
\end{equation}
can be written as the Mellin convolution of two functions $f_1$ and
$f_2$. As a result, the definite integral (\ref{Marichev}) can be
transformed to an inverse Mellin transform,
\begin{equation}
  f(z)
  =
  \frac{1}{2\pi i}
  \int_{\mathcal{L}} 
  {\mathfrak{M}}_{f_1}(u)\,{\mathfrak{M}}_{f_2}(u)\, 
  z^{-u}\,{\text{d}}u
\label{MellinBarnes}
\end{equation}
The power of this approach is that, if the functions $f_1$ and $f_2$
are of the hypergeometric type, which is true for many elementary
functions and the majority of special functions, the integral
(\ref{MellinBarnes}) turns out to be a Mellin-Barnes
integral. Depending on the involved coefficients, this integral can be
evaluated as a Fox $H$ function, or in simpler cases, a Meijer $G$
function (see Appendix~{\ref{Meijer.sec}}). 

The form of the equation~(\ref{depint}) allows to apply the Mellin
tranform method, with $z=1$ and
\begin{equation}
  f_1(t)
  =
  \frac{I_0}{R_{\text{e}}}\,
  \frac{b}{m\,\pi}\,
  {\text{e}}^{-bt^{1/m}} t^{\frac{1}{m}}
\end{equation}
and
\begin{equation}
  f_2(t)
  =
  \begin{cases}
    \;\dfrac{t}{\sqrt{1-s^2t^2}}
    &\qquad\text{if }0\leq t \leq s^{-1} \\
    \; 0
    &\qquad\text{else}
  \end{cases}
\end{equation}
The Mellin transforms of these functions are readily calculated
\begin{gather}
  {\mathfrak{M}}_{f_1}(u)
  =
  \frac{I_0}{R_{\text{e}}}\,
  \frac{1}{\pi}\,
  \frac{\Gamma(1+m\,u)}{b^{mu}}
 \\
  {\mathfrak{M}}_{f_2}(u)
  =
  \frac{\sqrt{\pi}\,\Gamma\left(\frac{1+u}{2}\right)}
  {\Gamma\left(\frac{u}{2}\right)}\,\frac{1}{u\,s^{1+u}}
\end{gather}
Substituting these values in the integral~(\ref{MellinBarnes}) and
setting $u=2x$, we obtain
\begin{equation}
 \nu(r)
  =
  \frac{I_0}{R_{\text{e}}}\,
  \frac{1}{\sqrt{\pi}}\,s^{-1}\,
 \frac{1}{2\pi i}
  \int_{\mathcal{L}} 
  \frac{\Gamma(1+2m\,x)\,\Gamma\left(\frac12+x\right)}
  {\Gamma(1+x)}\,
  (b^{2m}s^2)^{-x}\,
 {\text{d}}x
\label{nuHint}
\end{equation}
If we compare this expression with the definition~(\ref{defH}) of the
Fox $H$ function, we see that we can write the luminosity density of
the S\'ersic models in compact form as
\begin{equation}
  \nu(r)
  =
  \frac{I_0}{R_{\text{e}}}\,
  \frac{1}{\sqrt{\pi}}\,s^{-1}\,
  H^{2,0}_{1,2}\!\left[\left.
    \begin{matrix} 
      (1,1)\\
      (1,2m), (\tfrac12,1)
    \end{matrix}
    \,\right|\,
    b^{2m}s^2
  \right]
\label{nuH}
\end{equation}

\section{Integer and rational values of the S\'ersic index}
\label{DensityG.sec}

Expression~(\ref{nuH}) represents a closed, analytical expression for
the luminosity density of the S\'ersic model in terms of the Fox $H$
function. While this function is receiving gradually more attention
both in mathematics and applied sciences, ranging from astrophysics
and earth sciences to statistics, its practical usefulness is still
limited. In particular, no general numerical implementations of the
Fox $H$ function are available (to our knowledge). Fortunately, the
Fox $H$ function can be reduced to a Meijer $G$ function in many
cases. In this Section we will derive an analytical expression for the
luminosity density of the S\'ersic model in terms of the Meijer $G$
function for all rational values of $m$. The Meijer $G$ function is
much better documented and can be an extremely useful tool for
analytical work. The Meijer $G$ function has many general properties
which allow to manipulate expressions to equivalent forms, reduce the
order for certain values of the parameters, etc. Some of the most
useful properties are listed in Appendix~{\ref{Meijer.sec}}, but there
are many more. A particularly rich online source of information is the
Wolfram Functions Site.\footnote{The Wolfram Functions Site
  (\href{http://functions.wolfram.com/}{http://functions.wolfram.com/})
  is a comprehensive online compendium, providing a huge collection of
  formulas and graphics about mathematical functions. It is created
  with Mathematica$^\circledR$ and is developed and maintained by
  Wolfram Research with partial support from the National Science
  Foundation. A compendium of formulae on the Meijer $G$ function can
  be found at
  \href{http://functions.wolfram.com/PDF/MeijerG.pdf}{http://functions.wolfram.com/PDF/MeijerG.pdf}.}
Moreover, commercial computer algebra systems such as
Maple$^\circledR$ and Mathematica$^\circledR$ contain an
implementation of the Meijer $G$ function. It is also implemented in
the open-source computer algebra package Sage and a freely available
Python implementation of the Meijer $G$ function to arbitrary
precision is available from the Mpmath library.\footnote{Mpmath
  (\href{http://code.google.com/p/mpmath/}{http://code.google.com/p/mpmath/})
  is a free pure-Python library for multiprecision floating-point
  arithmetic. It provides an extensive set of transcendental
  functions, unlimited exponent sizes, complex numbers, interval
  arithmetic, numerical integration and differentiation, root-finding,
  linear algebra, and much more.}

\subsection{Integer and half-integer values of $m$}

If $m$ is an integer or half-integer value, we can simplify
expression~(\ref{nuH}) by using the property
\begin{equation}
  \prod_{j=0}^{N-1}
  \Gamma\!\left(\frac{j+z}{N}\right)
  =
  N^{\frac12-z}\,(2\pi)^{\frac{N-1}{2}}\,\Gamma(z)
  \label{prodGamma}
\end{equation}
Applying this recipe with $N=2m$ and $z=2m\,x$ gives
\begin{multline}
  \Gamma(1+2m\,x)
  =
  (2m)^{\frac12+2mx}\,(2\pi)^{\frac12-m}\,
  \\
  \times
  \Gamma(1+x)\,\prod_{j=1}^{2m-1}\Gamma\!\left(\frac{j}{2m}+x\right)
\end{multline}
Inserting this in expression~(\ref{nuHint}), we find
\begin{multline}
  \nu(r)
  =
  \frac{2\,I_0}{R_{\text{e}}}\,
  \frac{\sqrt{m}}{(2\pi)^m}\,s^{-1}\,
 \frac{1}{2\pi i}
  \\
 \times
  \int_{\mathcal{L}} 
  \Gamma\!\left(\frac12+x\right)
  \prod_{j=1}^{2m-1}\Gamma\!\left(\frac{j}{2m}+x\right)
  \left[
    \left(\frac{b}{2m}\right)^{2m}s^2
  \right]^{-x}\,
  {\text{d}}x
\end{multline}
Comparing this expression with the definition~(\ref{MeijerG}) of the
Meijer $G$ function, we obtain the following compact expression for
the luminosity density of the S\'ersic model
\begin{subequations}
\label{nuint}
\begin{equation}
  \nu(r)
  =
 \frac{2\,I_0}{R_{\text{e}}}\,
 \frac{\sqrt{m}}{(2\pi)^m}\,
 s^{-1}\,
  G_{0,2m}^{2m,0}
  \left[
    \begin{matrix} - \\ {\boldsymbol{b}} \end{matrix}
    \left|\,
    \left(\frac{b}{2m}\right)^{2m} s^2
    \right.
  \right]
\end{equation}
with ${\boldsymbol{b}}$ a vector with $2m$ elements given by
\begin{equation}
  {\boldsymbol{b}}
  =
  \left\{\frac{1}{2m},\frac{2}{2m}\ldots,
\frac{2m-1}{2m},
\frac12\right\}
\end{equation}
\end{subequations} 
This expression is equivalent to (and actually even slightly simpler
than) the expression obtained by \citet{2002A&A...383..384M} using the
computer algebra package Mathematica$^\circledR$. Notice that
\citet{2002A&A...383..384M} obtained their formula for integer values
of $m$ only, whereas our analysis shows that exactly the same
expression also holds for half-integer values of $m$.

\subsection{Rational values of $m$}

Interestingly, these results can be generalized for all
{\em{rational}} values of $m$. Setting $m=p/q$ with $p$ and $q$
integer numbers, we can write expression~(\ref{nuHint}) as
\begin{multline}
  \nu(r)
  =
  \frac{I_0}{R_{\text{e}}}\,
  \frac{1}{\sqrt{\pi}}\,s^{-1}\,
  \\
  \times
\frac{1}{2\pi i}
  \int_{\mathcal{L}} 
  \frac{q\,\Gamma(1+2p\,y)\,\Gamma\left(\frac12+q\,y\right)}
  {\Gamma(1+q\,y)}\,
  (b^{2p}s^{2q})^{-y}\,
  {\text{d}}y
\end{multline}
By multiple application of the identity~(\ref{prodGamma}) we can
rewrite this expression in a format that leads to a Meijer $G$
function. The result is
\begin{subequations}
\label{nugen}
\begin{equation}
  \nu(r)
  =
 \frac{2\,I_0}{R_{\text{e}}}\,
 \frac{\sqrt{p\,q}}{(2\pi)^p}\,
 s^{-1}\,
 G_{q-1,2p+q-1}^{2p+q-1,0}
  \left[
    \begin{matrix} {\boldsymbol{a}} \\ {\boldsymbol{b}} \end{matrix}
    \left|\,
    \left(\frac{b}{2p}\right)^{2p} s^{2q}
   \right.
  \right]
\end{equation}
with ${\boldsymbol{a}}$ a vector of dimension $q-1$ with elements
\begin{equation}
  {\boldsymbol{a}}
  =
  \left\{
  \frac{1}{q},\frac2q,\ldots,\frac{q-1}{q}
\right\}
\end{equation}
and ${\boldsymbol{b}}$ a vector with $2p+q-1$ elements given by 
\begin{equation}
  {\boldsymbol{b}}
  =
  \left\{
    \frac{1}{2p},\frac{2}{2p},\ldots,\frac{2p-1}{2p},
    \frac{1}{2q},\frac{3}{2q},\ldots,\frac{2q-1}{2q}
    \right\}
\end{equation} 
\end{subequations} 
It is straightforward to check that the general expression
(\ref{nugen}) reduces to the expression~(\ref{nuint}) for integer
values of $m$. Using the order reduction formulae (\ref{schrap}) and
(\ref{orderreduction}) of the Meijer $G$ function, one can also
demonstrate the expression (\ref{nugen}) reduces to~(\ref{nuint}) for
half-integer values of $m$.

\subsection{Special cases: $m=1$ and $m=\frac12$}
\label{specialcases.sec}

Among the family of S\'ersic models, there are two well-known specific
cases for which the luminosity density can be calculated explicitly in
terms of elementary or special functions. The first of these two
models is the exponential model, corresponding to $m=1$. Exponential
models are often used to describe the surface brightness profiles of
dwarf elliptical galaxies \citep[e.g.][]{1983ApJ...266L..17F,
  1984AJ.....89...64B}. If we introduce the notation
$h=R_{\text{e}}/b$, we can write the surface brightness profile as
\begin{equation}
  I(R)
  =
  I_0\,{\text{e}}^{-R/h}
\end{equation}
If we deproject this surface brightness profile using the deprojection
formula (\ref{deprojection}), we recover the well-known result that
the luminosity density can be written in terms of a modified Bessel
function of the second kind,
\begin{equation}
  \nu(r)
  =
  \frac{I_0}{\pi\,h}\,K_0\left(\frac{r}{h}\right)
  \label{nu1}
\end{equation}
If we set $m=1$ in the expression~(\ref{nuint}), we obtain
\begin{equation}
  \nu(r)
  =
 \frac{I_0}{\pi\,R_{\text{e}}}\,
 s^{-1}\,
  G_{0,2}^{2,0}
  \left[
    \begin{matrix} - \\ \frac12,\frac12 \end{matrix}
    \left|\,
    \frac{b^2s^2}{4}
    \right.
  \right]
\end{equation}
Using formula~(\ref{G2002}), this expression reduces to the
expression~(\ref{nu1}).

Another interesting special case is $m=\frac12$, which corresponds to
a gaussian surface brightness profile. Such profiles do not correspond
to the observed surface brightness profiles of galaxies, but they are
very useful as components in multi-gaussian expansions: even with a
relatively modest set of gaussian components, realistic geometries can
accurately be reproduced \citep[e.g.][]{1994A&A...285..723E,
  1994A&A...285..739E, 2001ApJ...546..903D, 2002MNRAS.333..400C}. If
we introduce $\sigma=R_{\text{e}}/\sqrt{2b}$ and we use the total
luminosity instead of the effective intensity as a parameter, we can
write the surface brightness profile as
\begin{equation}
  I(R)
  =
  \frac{L}{2\pi\sigma^2} \exp\left(-\frac{R^2}{2\sigma^2}\right)
\label{Igauss}
\end{equation}
One of the key advantages of a multi-gaussian expansion of an observed
surface brightness profile is that the corresponding luminosity
density has a simple analytical form. Indeed, substituting
(\ref{Igauss}) into the deprojection formula~(\ref{deprojection}), one
can easily check that the deprojection of a gaussian distribution on
the sky is a gaussian distribution as well,
\begin{equation}
  \nu(r)
  =
  \frac{L}{(2\pi\sigma^2)^{3/2}}
  \exp\left(-\frac{r^2}{2\sigma^2}\right)
  \label{nu12}
\end{equation}
This result can also be found by setting $m=\frac12$ in
equation~(\ref{nuint}),
\begin{equation}
  \nu(r)
  =
 \frac{I_0}{\sqrt{\pi}\,R_{\text{e}}}\,
 s^{-1}\,
  G_{0,1}^{1,0}
  \left[
    \begin{matrix} - \\ \frac12 \end{matrix}
    \left|\,
    bs^2
    \right.
  \right]
\end{equation}
If we use equation~(\ref{G1001}), we easily recover
expression~(\ref{nu12}).

\section{The total luminosity}
\label{totalluminosity.sec}

The total luminosity of the S\'ersic model can be calculated by
integrating the intensity on the plane of the sky,
\begin{equation}
  L
  =
  2\pi\int_0^\infty I(R)\,R\,{\text{d}}R
\end{equation}
Inserting equation~(\ref{SersicI}) one readily finds
\begin{equation}
  L
  =
  \pi\,I_0\,R_{\text{e}}^2\,
  \frac{1}{b^{2m}}\,
  \Gamma(2m+1)
\label{L}
\end{equation}
As a sanity check on the formula (\ref{nugen}) and as a illustration
of the power of the Meijer $G$ function as an analytical tool, we can
also calculate the luminosity by integrating the luminosity density
$\nu(r)$ over the entire space,
\begin{equation}
  L
  =
  4\pi\int_0^\infty \nu(r)\,r^2\,{\text{d}}r
\end{equation}
Inserting equation~(\ref{nugen}) we obtain
\begin{equation}
  L
  =
  \frac{2\,I_0\,R_{\text{e}}^2}{(2\pi)^{p-1}}\,
  \sqrt{\frac{p}{q}}
 \int_0^\infty
  G_{q-1,2p+q-1}^{2p+q-1,0}
  \left[
    \begin{matrix} {\boldsymbol{a}} \\ {\boldsymbol{b}} \end{matrix}
    \left|\,
      \left(\frac{b}{2p}\right)^{2p} t
    \right.
  \right]
  t^{\frac1q-1}\,{\text{d}}t
\end{equation}
If we use the general property~(\ref{defintMeijer}) of the Meijer $G$
function, we can evaluate this integral as
\begin{equation}
  L
  =
  \frac{2\,I_0\,R_{\text{e}}^2}{(2\pi)^{p-1}}\,
  \sqrt{\frac{p}{q}}
  \left(\frac{2p}{b}\right)^{\frac{2p}{q}}
 \frac{
    \prod_{j=1}^{2p+q-1}
    \Gamma\!\left(\frac{1}{q}+b_j\right)
  }{
    \prod_{j=1}^{q-1}
    \Gamma\!\left(\frac{1}{q}+a_j\right)
  }
  \label{Lpro}
\end{equation}
The product in the denominator can be simplified to
\begin{align}
  \prod_{j=1}^{q-1}
  \Gamma\!\left(\frac{1}{q}+a_j\right)
  =
 \frac{
    \prod_{j=0}^{q-1}\Gamma\!\left(\frac{j+1}{q}\right)
  }{
    \Gamma\!\left(\frac{1}{q}\right)
}
  =
  \frac{q^{-\frac12}
    (2\pi)^{\frac{q-1}{2}}}{\Gamma\!\left(\frac{1}{q}\right)}
\label{onder}
\end{align}
where the last transition follows from the identity (\ref{prodGamma}).
Similarly, the product in the numerator of equation~(\ref{Lpro}) can
be simplified to
\begin{align}
  \prod_{j=1}^{2p+q-1}
  \Gamma\!\left(\frac{1}{q}+b_j\right)
  &=
  \frac{
    \prod_{j=0}^{2p-1}\Gamma\!\left(\frac{j+2p/q}{2p}\right)
    \prod_{j=0}^{q-1}\Gamma\!\left(\frac{j+3/2}{q}\right)
  }
  {\Gamma\!\left(\frac1q\right)}
  \nonumber
  \\
  &=
  \frac{\sqrt{\pi}\,(2\pi)^{p+\frac{q}{2}-1}(2p)^{\frac12-\frac{2p}{q}}}
  {2\,q}\,
  \frac{\Gamma\!\left(\frac{2p}{q}\right)}{\Gamma\!\left(\frac1q\right)}
\label{boven}
\end{align}
If we substitute (\ref{onder}) and (\ref{boven}) into the expression
(\ref{Lpro}), and we use $m=p/q$, we recover the expression~(\ref{L})
for the total luminosity of the S\'ersic model, as required.

\section{Asymptotic behaviour}
\label{Asymptotic.sec}

One of the most useful properties of the general
expression~(\ref{nugen}) is that it allows to elegantly determine the
asymptotic behaviour of the luminosity density of the S\'ersic model
at small and large radii. It is well-known
\citep[e.g.][]{1991A&A...249...99C} that the S\'ersic models have a
cusp for $m>1$ and a finite luminosity density core at $m<1$; this can
be seen immediately by evaluating the integral (\ref{depint}) for
$r=0$ (which converges only for $m<1$),
\begin{equation}
  \nu_0
  =
  \frac{I_0}{R_{\text{e}}}\,
  \frac{b}{m\,\pi}\,
  \int_0^\infty
  {\text{e}}^{-bt^{1/m}} t^{\frac1m-2}\,{\text{d}}t
  =
  \frac{I_0}{\pi R_{\text{e}}}\,b^m\,\Gamma(1-m)
  \label{nu0}
\end{equation}
For a more detailed discussion on the behaviour of the luminosity
density at small radii, we can use the asymptotic
expansion~(\ref{asymp}) of the Meijer $G$ function. In particular,
this equation shows that the lowest order term in the expansion is
determined by the smallest components $b_k$ in the vector
${\boldsymbol{b}}$. This depends on the value of the S\'ersic index
$m$.

For $m<\frac13$, the smallest component of the vector
${\boldsymbol{b}}$ is $b_{2p}=\frac{1}{2q}$ and the second-smallest is
$b_{2p+1}=\frac{3}{2q}$. After some algebra, which involves similar
techniques as applied in Section~{\ref{totalluminosity.sec}}, we find
the asymptotic expansion
\begin{subequations}
\label{nuasymp}
\begin{equation}
  \nu(r)
  \sim
  \frac{I_0\,b}{\pi R_{\text{e}}}
  \left[
    \frac{\Gamma(1-m)}{b^{1-m}}
    +
    \frac12\,\frac{\Gamma(1-3m)}{b^{1-3m}}\,\,s^2
  \right]
  \label{nuasymp0-13}
\end{equation}
For $m=\frac13$, the smallest component is still $b_{2p}\equiv b_2=\tfrac16$,
but now the two components $b_1$ and $b_{2p+1}\equiv b_3$ are both equal to
$\tfrac12$. In this case we cannot use the expansion
formula~(\ref{asymp}), since this formula is only valid if all
components of the vector ${\boldsymbol{b}}$ are different.  For
$m=\tfrac13$ we find the expansion
\begin{multline}
  \nu(r)
  \sim
  \frac{I_0\,b}{\pi R_{\text{e}}}
  \left[
    \frac{\Gamma(\frac23)}{b^{2/3}}
    +
    \frac{3}{2}\ln\!\left(\frac{1}{s}\right)s^2
  \right.
  \\
  \left.
    -
    \frac14\,(3+2\gamma+2\ln b-6\ln 2)\,s^2
  \right]
  \label{nuasymp13}
\end{multline}
where $\gamma\approx0.57721566$ is the Euler-Mascheroni constant.  If
$\tfrac13<m<1$, $b_{2p}=\frac{1}{2q}$ remains the smallest component
of the vector ${\boldsymbol{b}}$, but the second-smallest is now
$b_1=\frac{1}{2p}$. One finds
\begin{equation}
  \nu(r)
  \sim
  \frac{I_0\,b}{\pi R_{\text{e}}}
  \left[
    \frac{\Gamma(1-m)}{b^{1-m}}
    +
    \frac{\sqrt{\pi}}{2m}\,\frac{\Gamma\left(\frac12-\frac1{2m}\right)}
    {\Gamma\left(1-\frac{1}{2m}\right)}\,
    s^{\frac1m-1}
  \right]
  \label{nuasymp13-1}
\end{equation}
For $m=1$, we again have two equal components in the vector
${\boldsymbol{b}}$ and we cannot readily apply
formula~(\ref{asymp}). The asymptotic expansion for small $r$ now
reads
\begin{equation}
  \nu(r)
  \sim
  \frac{I_0\,b}{\pi R_{\text{e}}}
  \left[
    \ln\!\left(\frac{1}{s}\right)
    -
    (\gamma+\ln b-\ln2)
  \right]
\end{equation}
Finally, if $m>1$, the smallest component is $b_{1}=\frac{1}{2p}$,
which leads to 
\begin{equation}
  \nu(r)
  \sim
  \frac{I_0\,b}{R_{\text{e}}}\,
  \frac{1}{2m\sqrt{\pi}}\,\frac{\Gamma\left(\frac12-\frac1{2m}\right)}
  {\Gamma\left(1-\frac{1}{2m}\right)}\,
  s^{-(1-\frac1m)}
\end{equation}
\end{subequations}
The five different asymptotic expansions (\ref{nuasymp}) demonstrate
the different behaviour of the luminosity density at small radii,
depending on the value of the S\'ersic index $m$. For $m<1$ the
S\'ersic model has a finite density core with the central luminosity
density given by equation~(\ref{nu0}). At $m=1$ the model has a
logarithmic cusp, at $m>1$ we have a power-law cusp with logarithmic
slope $1-\frac{1}{m}$. In particular, the de Vaucouleurs model has a
luminosity density profile that behaves as $s^{-3/4}$ as small radii
\citep{1976AJ.....81..807Y, 1987A&A...175....1M}. Surprisingly, the
S\'ersic models with $m<\frac12$ do not have a monotonically
decreasing luminosity density profile with increasing radius. In the
expansions~(\ref{nuasymp0-13}) and (\ref{nuasymp13}), the first
non-constant term has a positive coefficient and hence the luminosity
density increases with increasing radius in the nuclear region. The
same accounts $\frac13<m<\frac12$, since the coefficient of the second
term in the expansion~(\ref{nuasymp13-1}) is positive for
$\frac13<m<\frac12$ and negative for $\frac12<m<1$.

At large radii, a single formula for the asymptotic expansion holds
for all S\'ersic indices,
\begin{equation}
  \nu(r)
  \sim
  \frac{I_0}{R_{\text{e}}}
  \sqrt{\frac{b}{2\pi m}}\,
  {\text{e}}^{-bs^{1/m}}\,
  \left(\frac{1}{s}\right)^{1-\frac{1}{2m}}
\end{equation}
in agreement with the result derived by \citet{1991A&A...249...99C}.

\section{Some other properties of the S\'ersic model}
\label{Derived.sec}

The analytical expression~(\ref{nugen}) for the luminosity density of
the S\'ersic models allows to express other properties of this family
analytically in terms of the Meijer $G$ function. The most important
ones are the cumulative luminosity and the gravitational potential.

For a spherically symmetric system, they cumulative luminosity $L(r)$
can be calculated as
\begin{equation}
  L(r) 
  =
  4\pi \int_0^r \rho(r')\,r'^2\,{\text{d}}r'
\label{defLr}
\end{equation}
After substitution of expression~(\ref{nugen}) in
equation~(\ref{intMeijer}) we find
\begin{equation}
  L(r)
  =
  \frac{2\,I_0\,R_{\text{e}}^2}{(2\pi)^{p-1}}\,
  \sqrt{\frac{p}{q}}\,
  s^2\,
  G^{2p+q-1,1}_{q,2p+q}
  \!\left[\left.
    \begin{matrix} 1-\frac{1}{q}, {\boldsymbol{a}} \\
      {\boldsymbol{b}}, -\frac{1}{q} \end{matrix}
     \,\right|\left(\frac{b}{2p}\right)^{2p} s^{2q}
 \right]
\end{equation}
which reduces to 
\begin{equation}
  L(r)
  =
  \frac{2\,I_0\,R_{\text{e}}^2\,\sqrt{m}}{(2\pi)^{m-1}}\,
  s^2\,
  G^{2m,1}_{1,2m+1}
  \!\left[\left.
    \begin{matrix} 0 \\
      {\boldsymbol{b}}, -1 \end{matrix}
     \,\right|\left(\frac{b}{2m}\right)^{2m} s^2
 \right]
\end{equation}
for integer or half-integer values of the S\'ersic index $m$. Again,
this expression is equivalent to the expression found by
\citet{2002A&A...383..384M}. The asymptotic expansion of the
cumulative luminosity for small $r$ can be found in the same way as we
did for the luminosity density in Section~{\ref{Asymptotic.sec}}. One
finds after some calculation for $m<1$
\begin{subequations}
\begin{equation}
  L(r) 
  \sim 
  \frac43\,I_0\,R_{\text{e}}^2\,b^m\,\Gamma(1-m)\,s^3
\end{equation}
For $m=1$ we obtain
\begin{equation}
  L(r)
  \sim
  \frac{4\,I_0\,R_{\text{e}}^2\,b}{3}\,
  \left[
    \ln\left(\frac{1}{s}\right)
    +
    \left(\frac13-\gamma-\ln b + \ln 2\right)
  \right] s^3
\end{equation}
and for $m>1$  
\begin{equation}
  L(r) 
  \sim
  \frac{2\sqrt{\pi}\,I_0\,R_{\text{e}}^2\,b}{2m+1}\,
  \frac{\Gamma\left(\frac12-\frac1{2m}\right)}
  {\Gamma\left(1-\frac{1}{2m}\right)}\,
  s^{2+\frac1m}
\end{equation}
\end{subequations}
These asymptotic expressions can also be found by directly inserting
the equations~(\ref{nuasymp}) in formula~(\ref{defLr}).

If we assume that mass follows light (or in case the S\'ersic model is
used to describe the mass density), we can also calculate the
(positive) gravitational potential $\Psi(r)$. The most convenient way
in the present case is to use the formula
\begin{equation}
  \Psi(r)
  =
  G\,\Upsilon
  \int_r^\infty \frac{L(r')\,{\text{d}}r'}{r'^2}
\end{equation}
where the $\Upsilon$ is the mass-to-light ratio. The result reads
\begin{multline}
  \Psi(r)
  =
  \frac{G\,\Upsilon\,I_0\,R_{\text{e}}}{(2\pi)^{p-1}}\,
  \frac{\sqrt{p}}{q^{3/2}}\,
  s
  \\
  \times
  G^{2p+q,1}_{q+1,2p+q+1}
  \!\left[\left.
    \begin{matrix} 1-\frac{1}{q}, 1-\frac{1}{2q}, {\boldsymbol{a}} \\
      {\boldsymbol{b}}, -\frac{1}{2q}, -\frac{1}{q} \end{matrix}
     \,\right|\left(\frac{b}{2p}\right)^{2p} s^{2q}
 \right]
\end{multline}
or if we introduce the total mass $M=\Upsilon\,L$ using
expression~(\ref{L})
\begin{multline}
  \Psi(r)
  =
  \frac{GM}{R_{\text{e}}}\,
  \frac{b^{\frac{2p}{q}}}{(2\pi)^p\sqrt{pq}\,\,
    \Gamma\left(\frac{2p}{q}\right)}\,
  s
  \\
  \times
  G^{2p+q,1}_{q+1,2p+q+1}
  \!\left[\left.
      \begin{matrix} 1-\frac{1}{q}, 1-\frac{1}{2q}, {\boldsymbol{a}} \\
        {\boldsymbol{b}}, -\frac{1}{2q}, -\frac{1}{q} \end{matrix}
      \,\right|\left(\frac{b}{2p}\right)^{2p} s^{2q}
  \right]
\label{Psigen}
\end{multline}
For integer and half-integer values of the S\'ersic index $m$, this
expression simplifies to
\begin{multline}
  \Psi(r)
  =
  \frac{GM}{R_{\text{e}}}\,
  \frac{b^{2m}}{(2\pi)^m\sqrt{m}\,\,\Gamma(2m)}\,
 s
  \\
  \times
  G^{2m+1,1}_{2,2m+2}
  \!\left[\left.
    \begin{matrix} 0, \frac12 \\
      {\boldsymbol{b}}, -\frac12, -1 \end{matrix}
     \,\right|\left(\frac{b}{2m}\right)^{2m} s^2
 \right]
\end{multline}
This expression can be reduced slightly further since the coefficient
$\tfrac12$ appears in both the ${\boldsymbol{a}}$ and
${\boldsymbol{b}}$ coefficient vectors. Applying
equation~(\ref{schrap}), the final result reads
\begin{subequations}
\begin{equation}
  \Psi(r)
  =
  \frac{GM}{R_{\text{e}}}\,
  \frac{b^{2m}}{(2\pi)^m\sqrt{m}\,\,\Gamma(2m)}\,
 s
 \, 
 G^{2m,1}_{1,2m+1}
  \!\left[\left.
    \begin{matrix} 0 \\
      {\boldsymbol{b}}'
    \end{matrix}
     \,\right|\left(\frac{b}{2m}\right)^{2m} s^2
 \right]
\end{equation}
with ${\boldsymbol{b}}'$ a vector with $2m+1$ elements given by
\begin{equation}
  {\boldsymbol{b}}'
  =
  \left\{\frac{1}{2m},\frac{2}{2m}\ldots,
    \frac{2m-1}{2m},
    -\frac12,-1\right\}
\end{equation}
\end{subequations}
This expression is somewhat simpler than, but equivalent to
expression (28) in \citet{2002A&A...383..384M}.
 
Since the luminosity density of the S\'ersic models never falls
steeper than $r^{-1}$ at small radii, it is no surprise that all
S\'ersic models have a finite potential well for all values of
$m$. \citet{1991A&A...249...99C} already derived an expression for the
depth of the potential well using the general expression
\begin{equation}
  \Psi_0
  =
  -4G\,\Upsilon\int_0^\infty 
  \frac{{\text{d}}I}{{\text{d}}R}(R)\,R\,{\text{d}}R
\end{equation}
Applied to the S\'ersic model surface brightness profile, the result
reads [his equation (12)]
\begin{equation}
  \Psi_0
  =
  4G\,\Upsilon\,I_0\,R_{\text{e}}\,\frac{\Gamma(m+1)}{b^m}
\label{Psi0}
\end{equation}
Taking the limit $r\rightarrow0$ for the expression (\ref{Psigen}), we
find 
\begin{equation}
  \Psi_0
  =
  \frac{GM}{R_{\text{e}}}\,
  \frac{2b^{m}}{\pi}\,
  \frac{\Gamma(m)}{\Gamma(2m)}
\end{equation}
equivalent to (\ref{Psi0}). Using the expansion formulae for the
Meijer $G$ function, we can calculate the asymptotic expansion for the
potential at small radii. Not surprisingly, we again obtain different
expansions for $m$ smaller than, equal to, and larger than 1. After a
lengthly calculation, one finds for $m<1$ a quadratically decreasing
potential,
\begin{equation}
  \Psi(r)
  \sim
  \frac{GM}{R_{\text{e}}}
  \left[
    \frac{2b^{m}}{\pi}\,
    \frac{\Gamma(m)}{\Gamma(2m)}
    -
    \frac{b^{3m}}{3\pi m}\,\frac{\Gamma(1-m)}{\Gamma(2m)}\,s^2
  \right]
\end{equation}
For the exponential model $m=1$, one obtains
\begin{multline}
  \Psi(r)
  \sim
  \frac{GM}{R_{\text{e}}}
  \left\{
    \frac{2b}{\pi}
    -
    \frac{b^3}{3\pi}
    \left[
      \log\left(\frac{1}{s}\right)
    \right.
  \right.
  \\
  \left.
    \left.
      +
      \left(\frac12 - \gamma + \ln b - \ln 2 
        -\frac12\psi_{3/2}
        +\frac12\psi_{5/2}
      \right)
    \right]
    s^2
  \right\}
\end{multline}
where $\psi_z$ is the digamma function. For $m>1$ the potential
decreases more softly than quadratically,
\begin{multline}
  \Psi(r)
  \sim
  \frac{GM}{R_{\text{e}}}
  \left[
    \frac{2b^{m}}{\pi}\,
    \frac{\Gamma(m)}{\Gamma(2m)}
  \right.
  \\
  \left.
    -
    \frac{b^{2m+1}}{\sqrt{\pi}\,(m+1)\,(2m+1)}\,
    \frac{\Gamma\left(\frac12-\frac{1}{2m}\right)}
    {\Gamma\left(1-\frac{1}{2m}\right)\Gamma(2m)}\,
    s^{1+\frac{1}{m}}
  \right]
\end{multline}
Finally, at large radii, the potential of all S\'ersic models falls
off as
\begin{equation}
  \Psi(r) 
  \sim
  \frac{GM}{r}
\end{equation}
as required for a system with a finite mass.

\section{Conclusions}
\label{Summary.sec}

We have used the Mellin transform technique to derive a closed,
analytical expression for the spatial luminosity density $\nu(r)$ of
the S\'ersic model. For general values of the S\'ersic parameter $m$,
this expression is a Fox $H$ function. We derive simplified
expressions for $\nu(r)$ in terms of the Meijer $G$ function for all
rational values of $m$; for integer values of $m$ our results are
equivalent to the expressions found by
\citet{2002A&A...383..384M}. Our analytical calculations complement
other theoretical studies of the S\'ersic model
\citep{1991A&A...249...99C, 1997A&A...321..724C, 1999A&A...352..447C,
  2001MNRAS.326..869T, 2004A&A...415..839C, 2005PASA...22..118G,
  2007JCAP...07..006E} and, given the extended literature on the
analytical properties of the Meijer $G$ function, can be used to
further examine the properties of this model analytically. We have
investigated the asymptotic behaviour of the luminosity density at
small and large radii, and find a rich variety in behviour depending
on the value of $m$. We have also derived analytical expression for
derived quantities, in particular the cumulative luminosity and the
gravitational potential. Our results can also be used in practical
calculations: as implementations of the Meijer $G$ function are
nowadays available both in symbolic computer algebra packages and as
high-performance computing code, our results open up the possibility
to calculate the luminosity density of the S\'ersic models to
arbitrary precision.

\appendix
\onecolumn
\section{The Meijer $G$ and Fox $H$ functions}
\label{Meijer.sec}

The Meijer $G$ function is a universal, analytical function that has
been introduced as a generalization of the hypergeometric series by
\citet{Meijer36}. Nowadays it is more commonly defined as an inverse
Mellin transform, i.e.\ a path integral in the complex plane,
\begin{equation}
  G^{m,n}_{p,q}\!\left[\left.
      \begin{matrix} 
      {\boldsymbol{a}}\\{\boldsymbol{b}}
    \end{matrix}
    \,\right|\,
    z
  \right]
  \equiv
  G^{m,n}_{p,q}\!\left[\left.
    \begin{matrix} 
      a_1,\ldots,a_p\\b_1,\ldots,b_q
    \end{matrix}
    \,\right|\,
    z
  \right]
  =
  \frac{1}{2\pi i}
  \int_{\cal{L}}
  \frac{\prod_{j=1}^m \Gamma(b_j+s)\,\prod_{j=1}^n \Gamma(1-a_j-s)}
  {\prod_{j=m+1}^q \Gamma(1-b_j-s)\,\prod_{j=n+1}^p
    \Gamma(a_j+s)} \,
  z^{-s}\,{\text{d}}s
  \label{MeijerG}
\end{equation}
with ${\cal{L}}$ a path in the complex plane and ${\boldsymbol{a}}$
and ${\boldsymbol{b}}$ are two vectors of dimension $p$ and $q$
respectively. 

The general Meijer $G$ function can reproduce many commonly used
special functions, including Bessel functions, elliptic integrals and
hypergeometric functions. Some of the special cases are
\begin{gather}
  G^{1,0}_{0,1}\!\left[\left.
      \begin{matrix}-\\b\end{matrix}
    \,\right|\,z\right]
  =
  {\text{e}}^{-z}\,z^b
  \label{G1001}
\\
  G^{2,0}_{0,2}\!\left[\left.
      \begin{matrix}-\\b_1,b_2\end{matrix}\,\right|\,z\right]
  =
  2\,z^{\frac12(b_1+b_2)}\,K_{b_1-b_2}\left(2\sqrt{z}\right)
  \label{G2002}
\end{gather}
where $K_\nu(z)$ is the modified Bessel function of the second kind.

The Meijer $G$ function has numerous useful properties that allow to
transform expressions to equivalent expressions. For example, one can shift
all parameters by a given number,
\begin{equation}
  G^{m,n}_{p,q}
  \!\left[\left.
      \begin{matrix} a_1,\ldots,a_p\\ b_1,\ldots,b_q \end{matrix}
     \,\right|\,z\right]
 =
  z^{-\alpha}\,
  G^{m,n}_{p,q}
  \!\left[\left.
      \begin{matrix} a_1+\alpha,\ldots,a_p+\alpha\\ 
        b_1+\alpha,\ldots,b_q+\alpha \end{matrix}
   \,\right|\,z\right]
\label{shift}
\end{equation}
An important property is that one can reduce the order of the Meijer
$G$ function if one parameter appears in both the upper and lower
parameter vectors (depending on the position). For example, if one of
the $a_k$ with $n<k\leq p$ equals one of the $b_j$ with $1\leq j\leq
m$, then
\begin{equation}
  G^{m,n}_{p,q}
  \!\left[\left.
      \begin{matrix} a_1,\ldots,a_p\\ b_1,\ldots,b_q \end{matrix}
      \,\right|\,z\right]
  =
  G^{m-1,n}_{p-1,q-1}
  \!\left[\left.
      \begin{matrix} a_1,\ldots,a_{k-1},a_{k+1},\ldots,a_p
        \\ b_1,\ldots,b_{j-1},b_{j+1},\ldots,b_q \end{matrix}
      \,\right|\,z\right]
  \label{schrap}
\end{equation}
Another powerful property that allows to reduce the order of the
Meijer $G$ function in certain cases is the following 
\begin{subequations}
\label{orderreduction}
\begin{equation}
  G^{m,n}_{p,q}
  \!\left[\left.
    \begin{matrix} {\boldsymbol{a}} \\ {\boldsymbol{b}} \end{matrix}
     \,\right|\,z\right]
 =
 \frac{k^{1+\nu+(p-q)/2}}{(2\pi)^{(k-1)\delta}}\,
 G^{km,kn}_{kp,kq}
 \!\left[\left.
     \begin{matrix}
       {\boldsymbol{a}}' \\ {\boldsymbol{b}}'
     \end{matrix}
     \,\right|\,\frac{z^k}{k^{k(q-p)}}\right]
 \label{mult}
\end{equation}
where $k$ is a positive integer number, $\delta$ and $\nu$ are defined
as
\begin{gather}
  \delta 
  = 
  m+n+\frac{p+q}{2},
  \qquad
  \nu 
  =
  \sum_{j=1}^{q}b_j - \sum_{j=1}^p a_j
\end{gather}
and the vectors ${\boldsymbol{a}}'$ and
${\boldsymbol{b}}'$ are defined as
\begin{gather}
  {\boldsymbol{a}}'
  =
  \left\{
    \frac{a_1}{k}, \frac{a_1+1}{k},\ldots,\frac{a_1+k-1}{k},
    \ldots
    \frac{a_p}{k}, \ldots,\frac{a_p+k-1}{k}
  \right\}
  \\
  {\boldsymbol{b}}'
  =
  \left\{
    \frac{b_1}{k}, \frac{b_1+1}{k},\ldots,\frac{b_1+k-1}{k},
    \ldots
    \frac{b_q}{k},\ldots,\frac{b_q+k-1}{k}
  \right\}
\end{gather}
\end{subequations}
One of the most powerful properties of the Meijer $G$ function as an
analytical tool is that several integrals involving Meijer functions
can be evaluated in terms of higher-order Meijer functions. For
example,
\begin{equation}
  \int
  G^{m,n}_{p,q}
  \!\left[\left.
      \begin{matrix} a_1,\ldots,a_p \\ b_1,\ldots,b_q \end{matrix}
      \,\right|\,zw\right]
  z^{\alpha-1}\,
  {\text{d}}z
  =
  z^{\alpha}
  G^{m,n+1}_{p+1,q+1}
  \!\left[\left.
      \begin{matrix} 1-\alpha,a_1,\ldots,a_p \\ b_1,\ldots,b_q,-\alpha \end{matrix}
      \,\right|\,zw\right]
  \label{intMeijer}
\end{equation}
The corresponding definite integral can be evaluated as follows
\begin{equation}
  \int_0^\infty 
  G^{m,n}_{p,q}
  \!\left[\left.
      \begin{matrix} a_1,\ldots,a_p \\ b_1,\ldots,b_q \end{matrix}
      \,\right|\,zw\right]
  z^{\alpha-1}\,
  {\text{d}}z
  =
  \frac{
    \prod_{j=1}^m \Gamma(b_j+\alpha)\,
    \prod_{j=1}^n \Gamma(1-a_j-\alpha)
  }
  {
    \prod_{j=n+1}^p \Gamma(a_j+\alpha)\,
    \prod_{j=m+1}^q \Gamma(1-b_j-\alpha)
  }
  w^{-\alpha}
  \label{defintMeijer}
\end{equation}
Another useful property is the asymptotic expansion of the Meijer
function for small $z$, in the case of $p\leq q$ and simple poles,
\begin{equation}
  G^{m,n}_{p,q}
  \!\left[\left.
      \begin{matrix} {\boldsymbol{a}} \\ {\boldsymbol{b}} \end{matrix}
      \,\right|\,z\right]
  =
  \sum_{k=1}^m
  \frac{\prod_{j=1,j\ne k}^m \Gamma(b_j-b_k)\,
    \prod_{j=1}^n \Gamma(1-a_j+b_k)}
  {\prod_{j=n+1}^p \Gamma(a_j-b_k)\, \prod_{j=m+1}^q
    \Gamma(1-b_j+b_k)}\,
  z^{b_k}
  \left[
    1
    +
    (-1)^{-m-n+p}
    \frac{\prod_{j=1}^p (1-a_j+b_k)}{\prod_{j=1}^q (1-b_j+b_k)}\,
    z
    +
    {\cal{O}}(z^2)
  \right]
  \label{asymp}
\end{equation}
For the asymptotic expansion at large $z$, one can use the identity
\begin{equation}
  G^{m,n}_{p,q}
  \!\left[\left.
      \begin{matrix} a_1,\ldots,a_p \\ b_1,\ldots,b_q \end{matrix}
      \,\right|\,z\right]
  =
  G^{n,m}_{q,p}
  \!\left[\left.
      \begin{matrix} 1-b_1,\ldots,1-b_q \\ 1-a_1,\ldots,1-a_p \end{matrix}
      \,\right|\,\frac{1}{z}\right]  
\end{equation}

The Fox $H$ function \citep{Fox61} is a generalization of Meijer $G$
function and is also defined as an inverse Mellin transform,
\begin{equation}
  H^{m,n}_{p,q}\!\left[\left.
    \begin{matrix} 
      ({\boldsymbol{a}},{\boldsymbol{A}})
      \\
      ({\boldsymbol{b}},{\boldsymbol{B}})
    \end{matrix}
    \,\right|\,
    z
  \right]
  \equiv
  H^{m,n}_{p,q}\!\left[\left.
    \begin{matrix} 
      (a_1,A_1),\ldots,(a_p,A_p)\\(b_1,B_1),\ldots,(b_q,B_q)
    \end{matrix}
    \,\right|\,
    z
  \right]
  =
  \frac{1}{2\pi i}
  \int_{\cal{L}}
  \frac{\prod_{j=1}^m \Gamma(b_j+B_js)\,\prod_{j=1}^n \Gamma(1-a_j-A_js)}
  {\prod_{j=m+1}^q \Gamma(1-b_j-B_js)\,\prod_{j=n+1}^p
    \Gamma(a_j+A_js)} \,
  z^{-s}\,{\text{d}}s
  \label{defH}
\end{equation}
Not surprisingly, the Fox $H$ function shares many of the properties
of the Meijer $G$ functions and complete volumes have been written
about its identities, asymptotic properties, expansion formule and
integral transforms \citep[e.g.][]{MS78,KilbasSaigo,M+09}.

\end{document}